\documentclass[letterpaper, 10pt, conference]{ieeeconf}  

\IEEEoverridecommandlockouts                              

\usepackage{graphicx,multirow}

\usepackage{atbegshi,picture}
\AtBeginShipoutNext{\AtBeginShipoutUpperLeft{%
  \put(\dimexpr\paperwidth-1cm\relax,-1.5cm){\makebox[0pt][r]{Accepted manuscript IEEE/EMBS Neural Engineering (NER) 2021}}%
}}

\usepackage[colorinlistoftodos]{todonotes}

\title{Model-Agnostic Meta-Learning for EEG Motor Imagery Decoding in Brain-Computer-Interfacing}

\author{Denghao Li$^{1}$, Pablo Ortega$^{2}$, Xiaoxi Wei$^{2}$ and Aldo Faisal$^{1,2}$
\thanks{Brain \& Behaviour Lab, $^{1}$Department of Bioengineering,
        $^{2}$Department of Computing, Imperial College London, 
        London SW7 2AZ, UK; Correspondence: aldo.faisal@imperial.ac.uk. We achknowledge funding from to PO and a UKRI Turing AI Fellowship to AAF.} 
}

\begin{document}

\maketitle
\thispagestyle{empty}
\pagestyle{empty}


\begin{abstract}
We introduce here   the idea of Meta Learning for training EEG BCI decoders. Meta Learning is a way of training machine learning systems so they learn to learn.
We apply here meta learning to a simple Deep Learning BCI architecture and compare it to transfer learning on the same architecture. 
Our Meta learning strategy operates by finding optimal parameters for the BCI decoder so that it can quickly generalise between different users and recording sessions -- thereby also generalising to new users or new sessions quickly.
We tested our algorithm on the Physionet EEG motor imagery dataset.
Our approach increased motor imagery classification accuracy between 60 to 80\%, outperforming other algorithms under the little-data condition. We believe that establishing the meta learning or learning-to-learn approach will help neural engineering and human interfacing with the challenges of quickly setting up decoders of neural signals to make them more suitable for daily-life. 
\end{abstract}


\section{INTRODUCTION}
EEG motor imagery decoding via machine learning has been widely applied in brain-computer-interfaces (BCIs).
Since its introduction into BCI a few years ago, deep learning has been shown to help improving the classification accuracy in EEG-BCI and move beyond the limits of conventional signal processing performance \cite{walker2015deep,yang2015use,kumar2016deep,schirrmeister2017deep}. This is  best explained because deep learning systems offer feature learning to discover the underlying simplicity in signals over hand-crafted feature detection.
The down side is that there is a substantial amount of data  required to achieve a good classification performance for each user on any given session, which is often infeasible in real-world application scenarios.  Moreover, a model trained in one session and subject  performs poorly on different one and even on new sessions of the same subject, further increasing the amount of time dedicated to stabilizing the BCI control.  Thus, transferring a pre-trained model to a new session often penalises the accuracy due to discrepancies in sensor positions, the users mental fatigue or differences across subjects \cite{ortega2018compact}.
The lack of model transferability across different users and sessions, thus, does not scale with increasing amount of EEG data available for similar or even the same BCI tasks. 

\begin{figure}[b]
    \centering
    \includegraphics[width=\columnwidth]{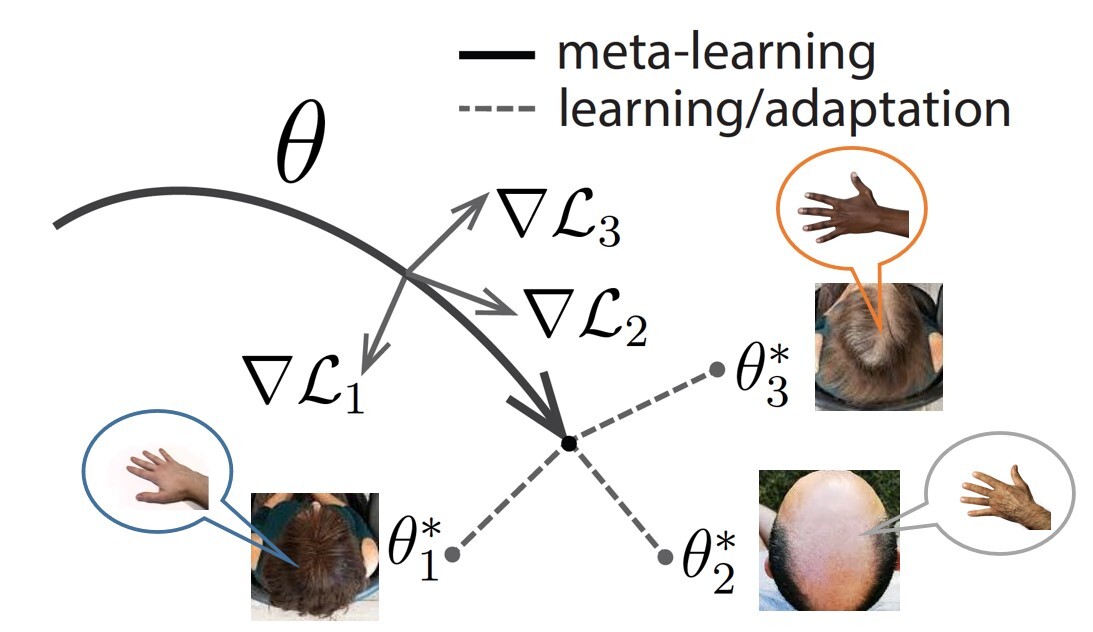}
    \caption{Model-Agnostic Meta-learning in EEG BCI learns and synthesises what is common between different users recordings and brains ("heads") as well as how they express their motor imagery ("right hand") to  transfers it to new user and their motor imagery. Figure adapted from \cite{finn2017model}.}
    \label{fig:2}
\end{figure}

One solution that has been proposed in machine learning based BCI decoding is the machine learning technique of transfer learning,
\cite{jayaram2016transfer,fahimi2019inter}. \emph{Transfer learning} (also referred to as \emph{domain adaptation} aims to help BCI decoders to generalise  to previous session or new users (in the context of neural engineering another "domain" corresponds here to data from other users or other sessions for the same user). Transfer learning methods differ from simply training BCI decoders with data from many users, in that they try to mitigate for the impact of negative transfer. Negative transfer is the result when combining data from different users or sessions actually decreases performance in a law of diminishing returns. This year, transfer learning  methods have been even successfully extended to deep learning BCI decoders and overcoming  negative transfer \cite{wei2021}. 
We are seeking here a different approach to transfer learning. In \emph{Meta Learning} we take, at least conceptually, a radically different approach, because we are focused on "learning to learn" \cite{vilalta2002perspective}. Meta learning as been shown to be effective in adapting models to learn new classification problems using only a few training points, making it data-efficient in contrast to transfer learning \cite{Ravi2017OptimizationAA,snell2017prototypical,munkhdalai2017meta}. We are introducing here the meta learning  approach to neural engineering, to address the challenge on how to learn to setup a machine learning system to learn the multi-user multi-session BCI decoding problem.  In other words, we allow our algorithm not only to learn a specific user's BCI decoding, but to learn the best way to learn the BCI decoding tasks in general. Having thus, data form multiple users and multiple sessions is thus an advantage as it meta learning benefits of having examples of different tasks (we treat here BCI decoding for different users and sessions as different tasks, as each requires its own personalised BCI decoder). The hope is that once our BCI meta learner knows how to learn the task, it requires less data to learn a new similar task. 
We focus here on model-agnostic meta-learning (MAML) \cite{finn2017model}, which does not assume a specific machine learning algorithms or deep learning architecture. In MAML, we will learn an initial set of parameters for a deep learning BCI decoder that can be fine-tuned easily on another similar task with only a few gradient steps (see Fig.~\ref{fig:2}).  
In MAML, deep learning BCI decoders perform gradient descent of the gradients associated to a task. However, the resulting second-order gradients required in MAML make it time and memory consuming. In practice, the first-order approximation of MAML (FOMAML) is applied more widely.

\section{METHODS}

Our methodological aim is to leverage the existing and novel data sets of EEG-BCI data to achieve better models in shorter times.  We develop meta-learning for our EEG-BCI classification settings, and regard each different subject as a different classification task. Additionally, we assume that the underlying brain function during motor imagery (the feature space) is  shared across subjects (the individual classification tasks) even if this is differently reflected in each subject's EEG signal.

\subsection{Models}
\begin{figure}[b]
    \centering
    \includegraphics[width=\columnwidth]{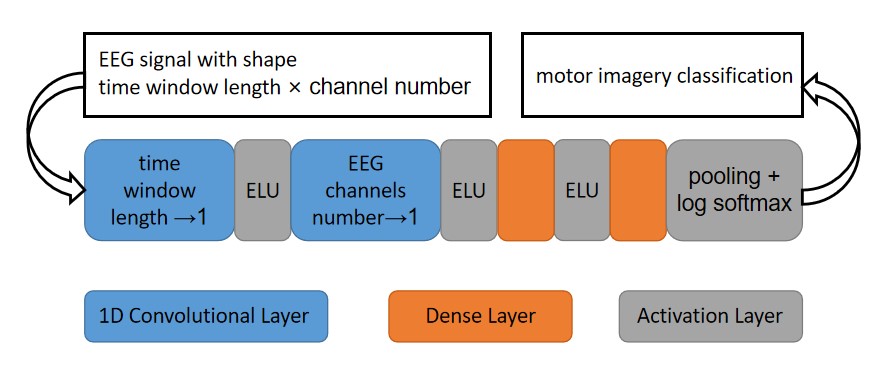}
    \caption{Neural network architecture.}
    \label{fig:1}
\end{figure}

Several models for EEG motor imagery classification have been developed in the past. In particular, common spatial patterns (CSP) is the most widespread algorithm for data-efficient, motor imagery classification (e.g. \cite{ferrante2015data}).
Since then Deep Learning methods have replaced these in academia (e.g. \cite{walker2015deep,yang2015use,kumar2016deep,schirrmeister2017deep}). It is even possible for each module in the deep learning architecture to reflect traditional signal processing steps \cite{ortega2021hemcnn}. We designed our meta-learning deep architecture with three processing stages (see Fig.~\ref{fig:1}).

In the first stage,  time domain oscillation patterns in the EEG signals are captured by the first layer via time-domain convolutions. These convolutions act as filters on all EEG channels along the time axis, extracting features that can be assimilated, but not limited, to the power of each frequency band. In particular, because the convolutional layer is entirely data-derived, it trains the filter bank to enhance the most active time-domain features across motor motor imagery examples. The second stage are spatial domain convolution, this corresponds to common spatial filters, that reveal the difference in activity intensity across different cortical areas. These first two stage are designed to preform feature learning.


Finally, in the third stage, the classification stage takes the output and maps through a two-layer convolutional neural network onto the BCI outputs. Because we map EEG time series through time domain convolutional we obtain a more compact time series of intermediate data. Therefore to ensure invariance to time translations, we average the output sequence in the time axis resulting in two outputs (hence the input sequence length is not limited in our model). This allows us to easily handle the trade-off between response time (proportional to temporal depth of our first layer's convolution) and prediction reliability, and thus we can experiment with different lengths in the input. Note, that when  Signal-to-Noise ratio of the raw data is high, the model can take short slices as inputs to give fast responses.  Conversely, longer inputs can be used to reduce the interference of noise with the cost of a longer delay.

\subsection{Training strategies}
The gist of our aim is fast adaptation to new subjects is to update the classification model with as little data.
The low signal-to-noise ration in EEG, means that training a model with little EEG data usually results up with unsatisfying BCI decoding performance, but we presume based on the neurobiology, that the fundamental EEG patterns of motor imagery are simple and generalisable between different people.

To evaluate the impact of meta-learning, we compare it against the only other model for data efficiency in training, namely transfer  learning.

\begin{figure}[htb]
    \centering
    \includegraphics[width=0.8\columnwidth]{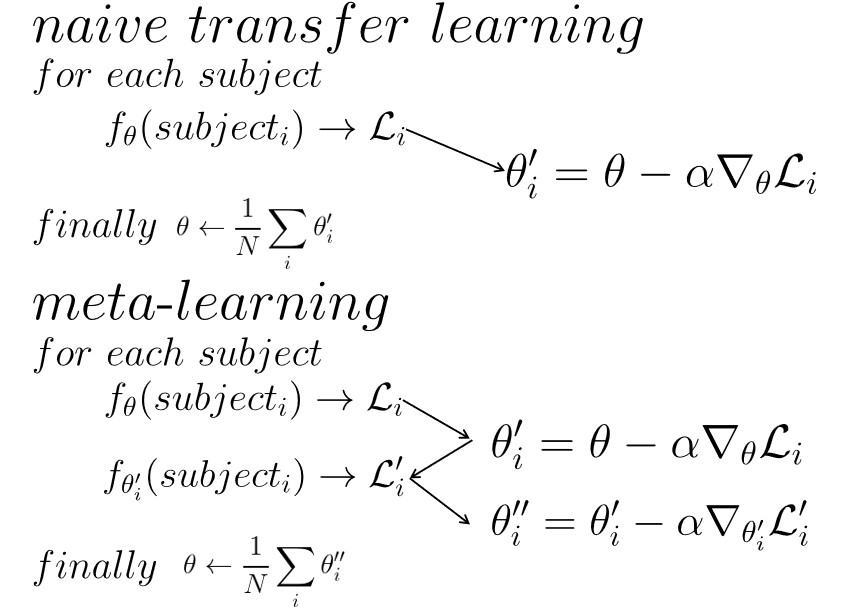}
    \caption{Algorithmic difference between transfer and meta learning in our BCI framework. See main text for details.}
    \label{fig:0}
\end{figure}

MAML shares similarities with  transfer learning in few-shot learning problems (see Fig.~\ref{fig:0}). In both cases, the exploration of a subject during training can be regarded as a loop of forward and backward steps.
Similar to the pre-training stage in naive transfer learning, the first step in MAML also aims to extract the general knowledge about the solution of the problem at hand.
In practice, this general knowledge consist of a pre-trained model, containing a set of general rules.
The model explores each problem (the classification of each subjects' EEG) with a step of gradient descent.
As a result, we obtain a set of gradients of the model parameters for the problem.
In naive transfer learning, the gradients from all subjects are effectively averaged before they update the model parameters.
However, averaging these gradients can cause important differences across subjects to cancel out leading to sub-optimal learning.
In contrast, in MAML each subject is seen as an independent sample of the same problem. Therefore, the gradients are computed for each subject separately. These subject-specific gradients are then backpropagated a set of subject-specific models is obtained. This model-set is in turn used to compute a new loss, which is minimised. Finally, the gradients of this new loss is used to update the parameters of the model.

Note, that if the loop over subjects stops after the first forward processing and $\mathcal{L}_{i}$ is minimized, it is naive transfer learning. If the loop stops after the second forward processing and $\mathcal{L}^{\prime}_{i}$ is minimized, it is MAML (pursued here). If the loop stops after more than two forward processing and $\mathcal{L}^{(n)}_{i}$ is minimized, it is another meta-learning algorithm which is known as ``reptile algorithm'' \cite{nichol2018first} (not pursued here). The meta-test stage of MAML is the same as the fine-tuning stage in naive transfer learning. The training set of the new subject to be tested is used to fine-tune the model.
If this stage is not based on a pre-trained model derived from transfer learning or MAML but a random initialization, it degenerates to conventional machine learning.

\subsection{Meta Learning with noisy EEG data}
The training process demonstrates how the models try to shape the distribution of the data. For a balanced binary EEG classification problem, the classification accuracy is empirically equivalent to the expectation that a sample in the dataset is correctly classified. This expectation can be regarded as a threshold to refine the decoded labels online.
Namely, only the samples with a predicted probability higher than this expectation should be accepted.
In contrast, those with a lower chance should be labelled as ``reject''. In practice, such filtering rejects most of the samples. However, in online motor imagery decoding applications, predictions could be obtained at a higher rate (e.g. every 300 milliseconds in our case) than the subject's may be able to operate. Therefore, it may be acceptable to sacrifice some of the response speed for the prediction accuracy.

\subsection{Data sets}
Motor imagery EEG BCI requires a suitable brain signals collection paradigm, reflecting on the fast responses during and simple setups. We used the motor imagery experiments in Physionet \cite{goldberger2000physiobank}.
This dataset includes four kinds of tasks: 
1) open and close left or right fist, 
2) imagine opening and closing left or right fist, 
3) open and close both fists or both feet, and 
4) imagine opening and closing both fists or both feet. 
We used tasks 1 and 4 and 17 channels covering the motor cortex (FCz $\sim$ FC4, Cz $\sim$ C6 and CPz $\sim$ CP4). 
Each motor imagery session was repeated for three times, and there were 15 trials in each session.
Each trial was about 4 seconds long  with rest between trials.
The total length of a session was less than 2 minutes. 
A bandpass filter from 8Hz to 45Hz was applied in the preprocessing. The data were then cropped in two-seconds windows for batch processing.

The dataset contains raw EEG data from 109 subjects and the data quality varies significantly across them.
Noise in the data can be so large that EEG signals cannot be recognised as such by eye. We there introduce here the 
the \emph{gambler's loss} \cite{ziyin2020learning} was optimized on each subject's data to identify outliers in terms of data quality.
Following this procedure, we finally selected 48 out of the 104 subjects for our experiments.
The probability of being an outlier was logged with each sample in the selected data for further analysis.

\section{RESULTS}


All computer experiments where conducted with the same deep learning network of the above section were conducted with the sample filtering procedure. We compared the performance of meta learning vs transfer learning strategies on the motor imagery tasks (Task 2 and 4) in the Physionet data as shown in Tab.~\ref{tab:5}. 
Predictions with low confidence level were filtered out by default.
In Meta-Learning the accuracy reached around 80\% after filtering nearly 80\% of the predictions (23.4\% and 23.0\% samples accepted for left fist vs right fist and both fists VS both feet respectively), shown in Tab.~\ref{tab:7}. In contrast  naive transfer learning failed to help the classification tasks on new subjects, indicating that indeed noise levels in EEG are high.

\begin{table}[t]
\centering
\caption{Test accuracy of different training strategies on Physionet data: left fist VS right fist (task 2); both fists VS both feet (task 4).}
\begin{tabular}{|c|cc|}\hline
    motor imagery task & task 2 & task 4 \\ \hline
    conventional learning & 59.6\% & 66.6\% \\
    transfer learning & 56.0\% & 62.4\% \\
    meta-learning & \textbf{64.5\%} & \textbf{68.2\%} \\
    \hline
\end{tabular}
\label{tab:5}
\end{table}

\begin{table}[b]
\centering
\caption{Test accuracy on Physionet data after online sample filtering}
\begin{tabular}{|c|cc|}
\hline
    motor imagery task & task 2 & task 4 \\ \hline
    without meta-learning & 67.9\% & 68.8\% \\
    with meta-learning & \textbf{80.6\%} & \textbf{79.7\%}\\
    \hline
\end{tabular}
\label{tab:7}
\end{table}

\begin{figure}[t]
    \includegraphics[width=0.5\textwidth]{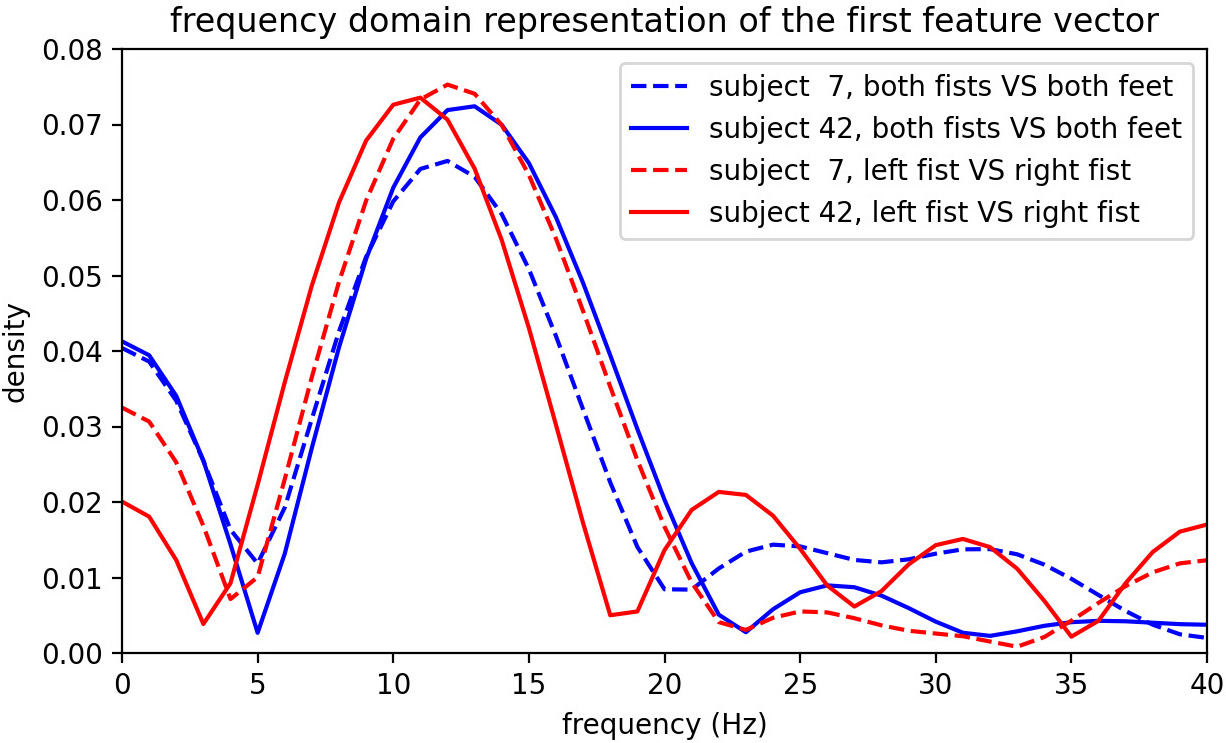}
    \caption{The frequency distributions of the singular vectors with the largest singular values in two example subjects. The colors of the curves represent different motor imagery tasks (blue for fists vs feet; red for left hand vs right hand). The solid and dashed lines indicate two typical subjects (solid line for subject 42 and dashed line for subject 7), we omit the other subjects to avoid crowding the plot.}
    \label{fig:5}
\end{figure}

We also opened the black-box of our deep learning network  to see what features it had learned in the first layer (where we can compute these analytically). We performed singular value decomposition  on the weight matrix of the first layer to extract the dominant component.
The largest singular value's component has a significant peak in the frequency domain, and the location of this peak is typically in the Alpha wave band, it thus appears that our network learned indeed something akin to time-domain filters (see Fig.~\ref{fig:5}).  


In summary, transfer and meta-learning training strategies were compared directly with each other to assess the performance of training an EEG BCI classifier on a new subject. We find that this first stab at meta-learning in BCI was shown to be a robust algorithm for the problem.

\section{DISCUSSION}

In this paper, we developed algorithms for EEG motor imagery classification with novel techniques in training strategy and data denoising.
We compared training strategies including conventional training, transfer learning and meta-learning.
The performance of transfer learning was even worse than conventional training.
The results reflected that the differences between subjects are too large for the model to capture the shared information in the group and optimize the loss synchronously among subjects.
On the other hand, meta-learning still brought improvements around 2\%.
In addition to the accuracy, meta-learning also cuts the time consuming of the online training processes.
Without meta-training, the model converges slowly after more than one hundred epochs.
However, with the information brought by meta-training, the model can converge ten times faster with the same optimizer and learning rate. For online EEG data denoising, training accuracy was used as a threshold for test data prediction.
The filtering approach improved the accuracy by about 10\% with a cost of about four times slower response. This performance compares well with the amount of training data needed to enable a single new user to be BCI decoded at \emph{par} performance as existing users.

The performance difference between meta-learning or transfer learning also reflects on underlying assumptions about the noise level in the EEG data. In transfer learning we pre-train a model using 
the data from a group of subjects, and fine-tune it with data 
from a new subject.  This transfer learning strategy overcomes the limitation in the amount of the  data during the training stage, given that the EEG patterns  evoked by the same imagery type can be classified into the same category across subjects. 
However, if differences are too small between  subjects, then these can be ignored, and pre-trained models 
could be directly used on new subjects or sessions.

Conversely, transfer learning should not be feasible when cross-subject differences are too large. Here meta-learning should operate better than transfer learning. 
\section{CONCLUSION}
In summary, the combination of our proposed algorithms leads to the improvement of the classification accuracy with a little training data from about 60\% to 80\% on the Physionet EEG motor imagery dataset.
The results show the potential to improve the feasibility of BCI systems in little-data scenarios, which will significantly reduce the burden on the calibration of  BCI use in daily-life applications.

\addtolength{\textheight}{-12cm}   








\bibliographystyle{IEEEtran}
\bibliography{IEEEabrv,mybibfile}

\end{document}